\documentclass[12pt]{article}
\usepackage[color]{healpix}

\input psfig

\begin{document}

\title{The \healpix Primer}
\docid{An Overview of the \healpix concept.}
\docrv{Version 1.0}
\author{Krzysztof M. G\'orski, Benjamin D.~Wandelt, Eric Hivon, Frode K.~Hansen,
and Anthony J. Banday}
\abstract{
\healpix is a Hierarchical, Equal Area, and iso-Latitude Pixelisation 
of the sphere  designed to support efficiently
(1) local operations on the pixel set,
(2) a hierarchical tree structure for  multi-resolution applications, and
(3) the global Fast Spherical Harmonic transform.
\healpix based mathematical software meets the challenges 
which future high resolution and large volume CMB data sets, 
including the MAP and Planck mission products, will present.
}
\date{\today}

\frontpage
\newpage 
\section{Introduction}

The analysis of functions on  domains with spherical topology occupies a
central place in physical science and engineering disciplines. 
This is particularly apparent in the fields of astronomy, cosmology, 
geophysics,  atomic and nuclear physics. In many cases the geometry is either
dictated by the object under study or approximate spherical symmetry can be
exploited to yield powerful perturbation methods. Practical
limits for the purely analytical study of these problems create
an urgent necessity for efficient and accurate numerical tools.

The 
simplicity of the spherical form belies the intricacy of global
analysis on the sphere. There is no known
point set which achieves the analogue of uniform sampling in Euclidean space and
allows exact and invertible discrete spherical harmonic decompositions
of arbitrary but band-limited functions. Any existing proposition of practical
schemes for the  discrete treatment of such functions 
on the sphere  introduces some (hopefully tiny) 
systematic error dependent on the global properties of
the point set. The goal is to minimise these errors and
faithfully represent deterministic functions as well as realizations of
random variates both
in configuration and Fourier space while maintaining computational efficiency.

We illustrate these points using as an example the field which is particularly 
close to the authors' hearts, Cosmic Microwave Background (CMB)
anisotropies. Here we are in the happy situation of expecting an explosion
of available data within the next decade.
The  Microwave Anisotropy Probe (MAP) (NASA) and Planck Surveyor (ESA)
missions are aiming
to provide multi-frequency, high resolution, full sky measurements of the anisotropy in
both temperature and polarization of the cosmic microwave
background radiation.
The ultimate data products of these missions ---
multiple microwave sky maps, each of which will have to comprise 
more than $\sim 10^6$ pixels in order to render the angular 
resolution of the instruments ---
will present serious challenges to those involved in the
analysis and scientific exploitation of the results of both surveys.
 
As we have learned while working with the {\it COBE} mission products, 
the digitised
sky map is an essential intermediate 
stage in information processing between 
the entry point of data acquisition by the 
instruments --- very large time ordered data streams,
and the final stage of astrophysical analysis --- 
typically producing a $\lq$few' numerical values
of physical parameters of interest. 
{\it COBE}-DMR sky maps (angular resolution of $7^\circ$ (FWHM) in
three frequency bands, two channels each, 6144 pixels per map)
were considered large at the time of their release.

As for future CMB maps, a whole sky CMB survey  
at the angular resolution
of $\sim 10'$ (FWHM), discretised with 
a few pixels per resolution element 
(so that the discretisation effects on the signal are  
sub-dominant with respect to the effects of instrument's angular response),
will require map sizes of at least 
$N_{pix}\sim $ a few $\times 1.5\, 10^6$ pixels.
More pixels than that will be needed to represent the Planck-HFI higher 
resolution channels.
This estimate, $N_{pix}$, should be multiplied by  the number of frequency bands 
(or, indeed, by the number of individual
observing channels --- 74 in the case of Planck --- for the analysis work 
to be done before the
final coadded maps are made for each frequency band) to render 
an approximate expected
size of the already very compressed form of survey data which would 
be the input to the astrophysical analysis pipeline. 

It appears to us that very careful
attention ought to be given to devising  high resolution CMB map
structures which can maximally facilitate 
the forthcoming analyses of large size data sets, for the following
reasons:
\begin{itemize}
\item It is clearly very easy to end up with an estimated size of many GBy 
for the 
data objects which would be directly involved in the science extraction
part of the future CMB missions.
\item Many essential scientific questions
can only be answered by {\em global} studies of  future data sets.
\end{itemize}

This document is an introduction to the
properties of our proposed approach for a high resolution numerical
representation of  functions on the sphere
 --- the Hierarchical
Equal Area and isoLatitutde Pixelization  (\healpixns, see 
{\tt http://www.tac.dk/$\sim$healpix}), and the associated multi-purpose computer 
software package. 
 
\section{Discretisation of Functions on the Sphere for 
High Resolution Applications:
a Motivation for \healpix}

Numerical analysis of functions on the sphere involves 
(1) a class of mathematical operations, whose objects are 
(2) discretised maps, i.e. quantizations of arbitrary functions 
according to a
chosen tessellation (exhaustive partition of the sphere into 
finite area elements). Hereafter we mostly specialise our discussion 
to CMB related applications of
\healpixns, 
but all our statements hold true generally for any relevant 
deterministic and random functions on the sphere.

Considering point (1):
Standard operations of numerical analysis which one might wish to
execute on the sphere include
convolutions with local and global kernels, 
Fourier analysis with spherical harmonics
and power spectrum estimation,
wavelet decomposition, nearest-neighbour searches, topological
analysis, including searches for extrema or zero-crossings, 
computing Minkowski functionals,
extraction of patches and
finite differencing for solving partial
differential equations.
Some of these operations become prohibitively slow  
if the sampling of functions on the sphere, and the related structure of 
the discrete data set, are not designed carefully. 

Regarding point (2): 
Typically, a whole sky map rendered by a CMB experiment contains 
({\it i}) signals coming from the sky,
which are by design strongly band-width limited (in the sense of 
spatial Fourier
decomposition) by the instrument's angular response 
function, and 
({\it ii}) a projection into the elements of a discrete map, or pixels,
of the observing instrument's noise; this pixel noise should be random,
and white, at least near the discretisation scale, with a band-width 
significantly exceeding that of all the signals. 

With these considerations in mind we propose the following list of 
desiderata 
for the mathematical structure of discretised full sky maps:

{\bf 1. Hierarchical structure of the data base}. This is recognised as 
essential for very large data bases, and was  postulated 
in construction
of the Quadrilateralized Spherical Cube 
(or quad-sphere, see {\tt http://www.gsfc.nasa.gov/ast\-ro/co\-be/skymap\_info.html}), 
which was used for the
{\it COBE} data. An  argument in favour of this 
proposition
states that the data elements  
which are nearby in a multi-dimensional configuration space 
(here, on the surface of 
a sphere), are also nearby in the tree structure of the data base, hence
the near-neighbour searches are conducted optimally in the data storage medium
or computer RAM.
This property, especially when implemented with small number of base
resolution elements, 
facilitates various topological methods of analysis, 
and allows easy construction
of wavelet transforms on quadrilateral (and also triangular) grids. 
Figure~1 shows how a hierarchical partition with 
quadrilateral structure naturally allows for a binary vector indexing
of the data base.

\begin{figure}[!t]
\centerline{\psfig{figure=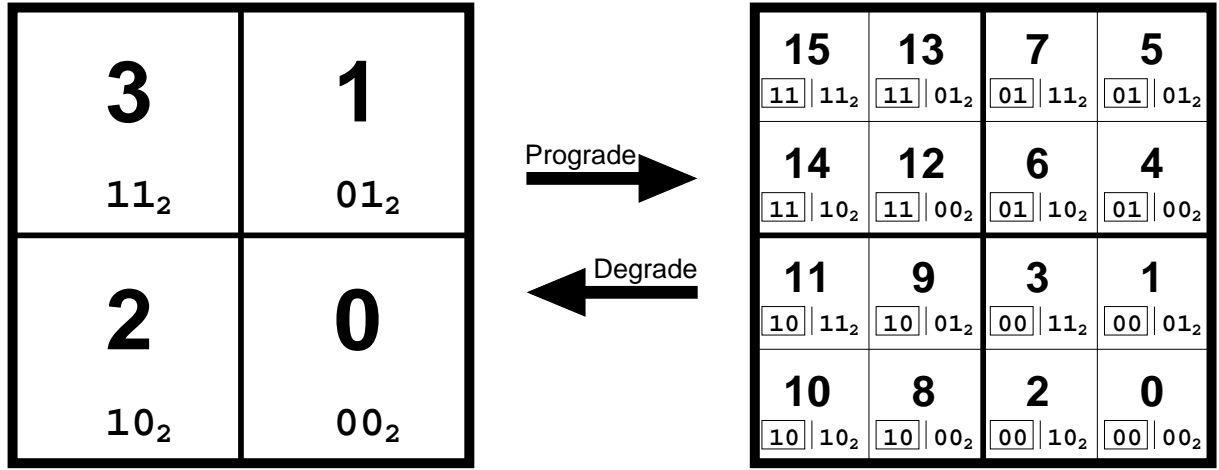,width=\textwidth}}
\caption
{Quadrilateral tree pixel numbering. 
The coarsely pixelised coordinate patch on
the left consists 
of four pixels. Two bits suffice to label the pixels. 
To increase the resolution, every 
pixel splits into 
4 daughter pixels shown on the right. These daughters inherit the pixel
index of their 
parent (boxed) and acquire 
two new bits to give the new pixel index. 
Several such curvilinearly mapped coordinate patches 
(12 in the case of \healpixns, and 6 in the case of the {\it COBE} quad-sphere) 
are joined at the boundaries to cover
the sphere. All pixels indices carry a prefix (here omitted for clarity) 
which identifies which base-resolution pixel they belong to.
\label{fig:quadtree}}
\end{figure}

{\bf 2. Equal areas of discrete elements of partition}. This is advantageous 
because ({\it i})
white noise generated by the  signal receiver 
gets integrated exactly into
white noise in the pixel space, and 
({\it ii}) sky signals are sampled without regional dependence, except for 
the dependence on pixel shapes, which is unavoidable with tessellations of the 
sphere. 
Hence, as much as possible given the experimental details, the pixel
size should be made sufficiently small compared to the 
instrument's resolution to avoid any excessive, and pixel shape dependent, 
signal smoothing.

{\bf 3. Iso-Latitude distribution of discrete area elements on a sphere}.  
This property
is critical for computational speed of all operations involving evaluation of 
spherical
harmonics. Since the associated Legendre polynomial components of
spherical harmonics are evaluated via
slow recursions, and 
can not be simply handled in an analogous way to the trigonometric Fast Fourier Transform, 
any deviations in the sampling grid from an iso-latitude
distribution result in a prohibitive loss of computational performance
with the growing number of sampling points, or increasing map resolution.
It is precisely this property that the {\it COBE} quad-sphere is lacking,
and this renders it impractical for applications to high resolution data.

A number of tessellations 
have been  used for discretisation and analysis 
of functions on the sphere (for example, see  
Driscoll \& Healy (1994),
Muciaccia, Natoli \& Vittorio (1998) --- rectangular grids,
Baumgardner \& Frederickson (1985),  Tegmark (1996) --- icosahedral grids,
Saff \& Kuijlaars (1997),  Crittenden \& Turok (1998) ---  `igloo' grids,
and Szalay \& Brunner (1998)  --- a triangular grid), but none 
satisfies simultaneously all three stated requirements.

All three requirements formulated above are satisfied by construction with the
Hierarchical Equal Area, iso-Latitude Pixelisation (\healpixns) 
of the sphere (G\'orski (1999)), which is shown in Figure~2.

\begin{figure}[!t]
\centering\mbox{\psfig{figure=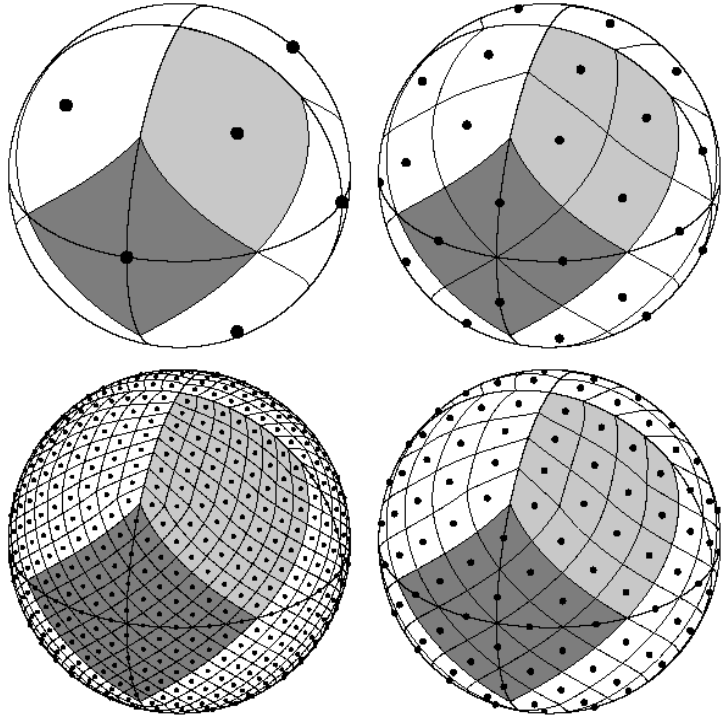,height=9.5cm}}
\caption[]{Orthographic view of \healpix partition of the sphere. 
Overplot of equator and  meridians illustrates the octahedral symmetry of  
\healpixns. 
Light-gray shading shows one of the eight (four north, and four south) 
identical polar 
base-resolution pixels. 
Dark-gray shading shows one of the four identical equatorial 
base-resolution pixels. 
Moving clockwise from the upper left 
panel the grid is hierarchically subdivided with 
the grid resolution parameter equal to $N_{side} = \,1,\,2,\,4,\,8$, 
and the total number of pixels  equal to 
$N_{pix} = 12 \times N_{side}^2 = \,12,\,48,\,192,\,768$. 
All pixel centers are located on $N_{ring} = 4 \times N_{side} - 1$ rings of 
constant latitude.
Within each panel the areas of all pixels are identical.}
\label{HEALPIX}
\end{figure}

\section{Geometric and Algebraic Properties of \healpix}

\healpix is a genuinely curvilinear partition of the sphere into exactly equal area
quadrilaterals of varying shape. The base-resolution comprises twelve pixels in three
rings around the poles and equator. 

The resolution of the grid is expressed by the parameter $N_{side}$ which defines the number
of divisions along the side of a base-resolution pixel that is needed to reach a desired
high-resolution partition.

All pixel centers are placed on $ 4\times N_{side}-1$ 
rings of constant latitude, 
and are equidistant in azimuth
(on each ring). All iso-latitude rings located between the upper and lower corners of
the equatorial base-resolution pixels, the equatorial zone, 
are divided into the same number of pixels: 
$N_{eq}= 4\times N_{side}$. The remaining rings are located within the
polar cap regions and contain a varying number of pixels, increasing 
from ring to ring with increasing distance
from the poles by one pixel within each quadrant. 

Pixel boundaries are non-geodesic and take the very simple 
forms $\cos \theta = a \pm b \cdot \phi$ in the equatorial zone, 
and $\cos \theta = a + b / \phi^2$, or
$\cos \theta = a + b / (\pi/2 - \phi) ^2$,
in the polar caps. 
This allows one to explicitly check by simple analytical integration the 
exact area equality among pixels,
and to compute efficiently more complex objects, 
e.g. the Fourier transforms of individual pixels.

\begin{figure} [!]
\centering\mbox{\psfig{figure=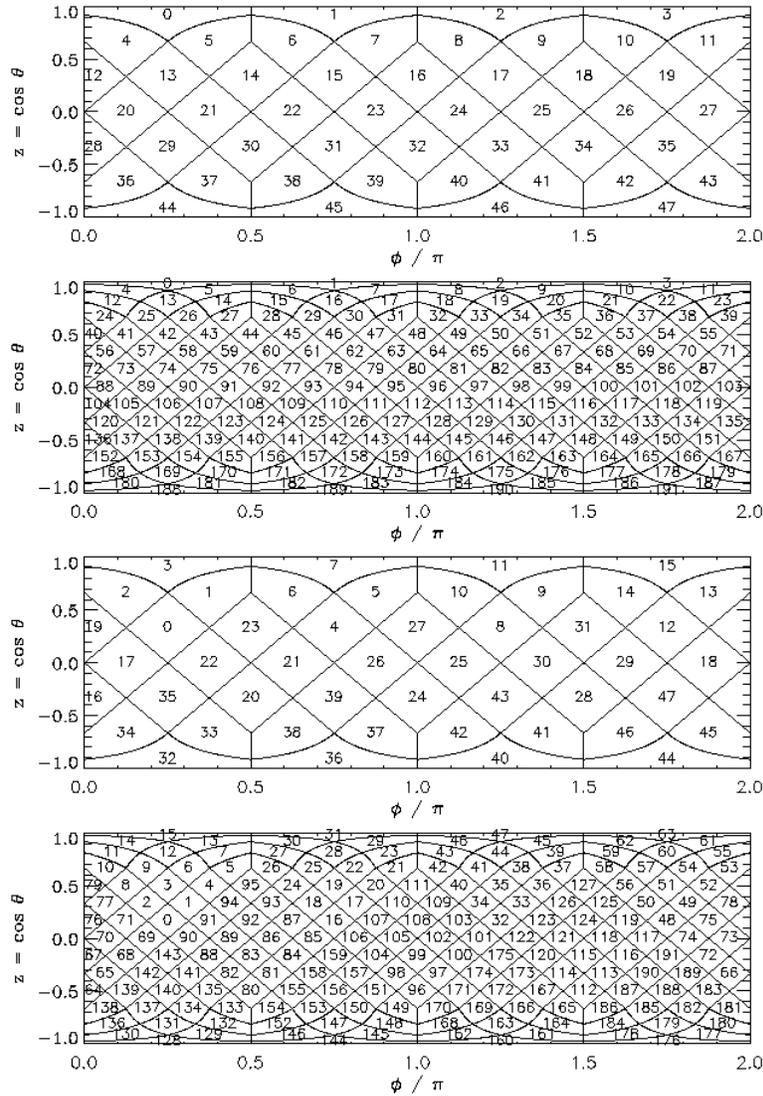,height=14.5cm}}
\caption[]{Cylindrical projection of the \healpix  division of a
sphere and two natural pixel numbering schemes (RING and NESTED) 
allowed by \healpixns. Both numbering schemes map the two dimensional 
distribution
of discrete area elements on a sphere into the one dimensional, 
integer pixel number array,
which is essential for computations involving data sets with very 
large total pixel numbers.
From top to bottom:
Panel one (resolution parameter $N_{side} = 2$) and panel two ($N_{side} = 4$)
show the RING scheme for pixel numbering, with the pixel number winding 
down from north to south pole through the consecutive isolatitude rings.
Panel three (resolution parameter $N_{side} = 2$) and panel four ($N_{side} = 4$)
show the NESTED scheme for pixel numbering within which the pixel number grows
with consecutive hierarchical subdivisions on a tree structure seeded by 
the twelve 
base-resolution pixels. 
}
\label{Numbering}
\end{figure}

Specific geometrical properties allow \healpix to support two different
numbering schemes for the pixels, as illustrated in the Figure~3.

First, in the RING scheme, 
one can simply count the pixels moving down from the north 
to the south pole along each
iso-latitude ring. It is in the RING scheme that Fourier transforms 
with spherical harmonics
are easy to implement.

Second, in the NESTED scheme, one can arrange the pixel indices 
in  twelve tree structures, corresponding to base-resolution pixels.
Each of those is organised as shown in Fig. 1.
This can easily be implemented
since, due to the simple 
description of pixel boundaries, the analytical mapping of the \healpix
base-resolution elements (curvilinear
quadrilaterals) into a [0,1]$\times$[0,1] square exists.
This tree structure allows one to implement efficiently all
applications involving  nearest-neighbour searches
(see Wandelt, Hivon \& G\'orski (1998)),
and also allows for an immediate
construction of the fast Haar wavelet transform on \healpixns.

\section{The \healpix Software Package}

We have developed a package of \healpix based mathematical software, consisting
of Fortran90 and IDL source codes as well as documentation and
examples. Successful installation produces a set of facilities using standardised
FITS I/O interfaces 
({\tt http://heasarc.gsfc.nasa.gov/docs/software/fitsio}) 
as well as two Fortran90
libraries  which users can link to their own applications.
Among the tasks performed by the  components of the 
\healpix package  are the
following:

\begin{itemize}
\item Simulation of the full sky CMB temperature and polarisation maps
as realisations of random Gaussian fields, with an option to constrain
the realisation by prior information.
\item Analysis of the full sky CMB temperature and polarisation maps
resulting in power spectra and/or spherical harmonic coefficients.  
\item Global smoothing of whole sky maps with a Gaussian kernel.
\item Degradation and upgrade of the resolution of discrete maps.
\item Global searches on the maps for nearest-neighbours and 
the maxima/minima of the discretised functions.
\item Algebraic conversion of the maps between the RING and NESTED numbering
schemes, and mapping back and forth between  positions on the sphere and 
discrete pixel index space.
\item Visualisation of the \healpix formatted sky maps in both the Mollweide 
and the gnomonic projections of  small areas of the sky. 
\end{itemize}

The package includes documents which describe the installation
process, the Fortran 90
facilities, the IDL facilities and a large number of 
subroutines contained in the library. It is
available to the scientific community  at {\tt
http://www.tac.dk/$\sim$healpix}. 

\healpix is  the format chosen by the MAP 
collaboration 
to be used for the production
of sky maps (see {\tt
http://map.gsfc.nasa.gov/html/technical\_info.html}) from the mission data. 
\healpix software is widely used 
for simulation work within both the LFI and HFI consortia of the Planck collaboration.


\newpage
\begin{thebibliography}{}

\bibitem{baum}
Baumgardner, J.R. and Frederickson, P.O., 1985, SIAM J. Numerical Analysis, Vol. 22,
No. 6, p. 1107

\bibitem{crtu}
Crittenden, R. and Turok, N.G., 1998, astro-ph/9806374

\bibitem{drhea}
Driscoll, J.R. and Healy, D., 1994, Adv. in Appl. Math., Vol. 15, p.202

\bibitem{gorhivwan}
G{\'o}rski K.M., Hivon, E. and  Wandelt, B.D., 1998, 
``Analysis Issues for Large CMB Data Sets'', astro-ph9812350,
to appear in Proceedings of the MPA/ESO Conferece on
Evolution of Large-Scale Structure: from Recombination to Garching
2-7 August 1998; eds. A.J. Banday, R.K. Sheth and L. Da Costa

\bibitem{ghealpix}
G{\'o}rski K.M., 1999, in preparation

\bibitem{munavi}
Mucaccia, P.F, Natoli, P. and Vittorio, N., 1998, Ap.J., 488, L63

\bibitem{mathint}
Saff, E.B. and Kuijlaars, A.B.J., 1997, The Mathematical 
Intelligencer, 19, \#1, p.5

\bibitem{szalay}
Szalay, A.S. and Brunner, R.J., 1998, astro-ph9812335, to appear in 
a special issue of the Elsevier journal "Future Generation Computer Systems"

\bibitem{teg}
Tegmark, M., 1996, 470, L81

\bibitem{whg}
Wandelt, B.D., Hivon, E. and G\'orski, K.M., 1998, astro-ph/9803317, in
"Fundamental Parameters in Cosmology", proceedings of the XXXIIIrd Rencontres
de Moriond, Tr$\hat{{\rm a}}$n Thanh V$\hat{{\rm a}}$n (ed.)
 
\end{thebibliography}
\end{document}